\documentclass[intlimits,twoside,a4paper]{article}

\usepackage[cp1251]{inputenc}

\usepackage[eqsecnum]{cmpj3}

\articletype{Rapid Communication}

\issue{2019}{22}{1}{14701}
\doinumber{10.5488/CMP.22.14701}

\title{Excited state properties of polycyclic hydrocarbons based dyes}
\author[Yu.V. Klysko, S.V. Syrotyuk]{Yu.V. Klysko, S.V. Syrotyuk}
\address{Semiconductor Electronics Department, Lviv Polytechnic National University, \\S. Bandera St., 12, 79013 Lviv, Ukraine
}

\date{Received February 27, 2019, in final form March 5, 2019}

\begin{document}

\maketitle

\begin{abstract}
In this paper we present a comprehensive ab initio study of polycyclic hydrocarbons based dyes. The purpose of the work is to obtain electronic properties of the materials which are supposed to be used in organic electronic devices. The list of materials includes violanthrone (di-benzanthrone) derivatives which are already known as industrial organic dyes. First, we have obtained ground-state properties by performing ab initio eigenvalue calculation within generalized gradient approximation (GGA). Then, Green's function method has been used in order to obtain excited state properties. The exciton eigenvalues, as well as imaginary part of dielectric function~(DF) and density of states (DOS), have been evaluated from the Bethe-Salpeter equation (BS). The electronic properties obtained here are in good agreement with available experimental data.
\keywords ab initio,  organic semiconductors, electronic structure
\pacs 71.20.Rv, 71.35.-y, 71.10.-w
\end{abstract}

\section{Introduction}

Recently, the interest to polyaromatic hydrocarbons has been growing due the rise of organic electronics and organic photovoltaics espesially, because it is supposed that planar molecules could replace fullerenes  \cite{Giant_Polycyclic_Aromatic_Hydrocarbons, doi:10.1021/ja000832x, 10.1351/PAC-CON-09-07-07, doi:10.1021/ja020323q, doi:10.1080/10406630701268255, C3TC32315C}. At present, perylene diimide (PDI) derivatives are the most popular materials due to their remarkable electronic properties, stability and processability \cite{zhan2011rylene, li2012perylene, C6QM00247A, C2CS15313K, C4MH00042K, yan2018non}. The rapid development of chemistry and investigations of these molecules has resulted in a high number of possible derivatives and growth of efficiency of the non-fullerene organic photovoltaics (OPV). Violanthrone has a structure similar to the structure of PDI. The history of violanthrone began in 1950, when its excellent, and similar to inorganic solids, electronic properties were investigated \cite{doi:10.1063/1.1747780, doi:10.1063/1.1700784}.  It consists of 9 benzene rings, so it has a larger $\piup$-conjugated system. Besides the use of violanthrones in OPV devices, it is also quite promising for organic thin film transistors \cite{SHI2012377, doi:10.1021/jp109683h} and light emitting devices, due to the emission in the red and infrared part (IR) of the spectrum  \cite{doi:10.1021/jz200270w, XIAO201532}.

\begin{figure}[!t]
\centering
\includegraphics[width=\linewidth]{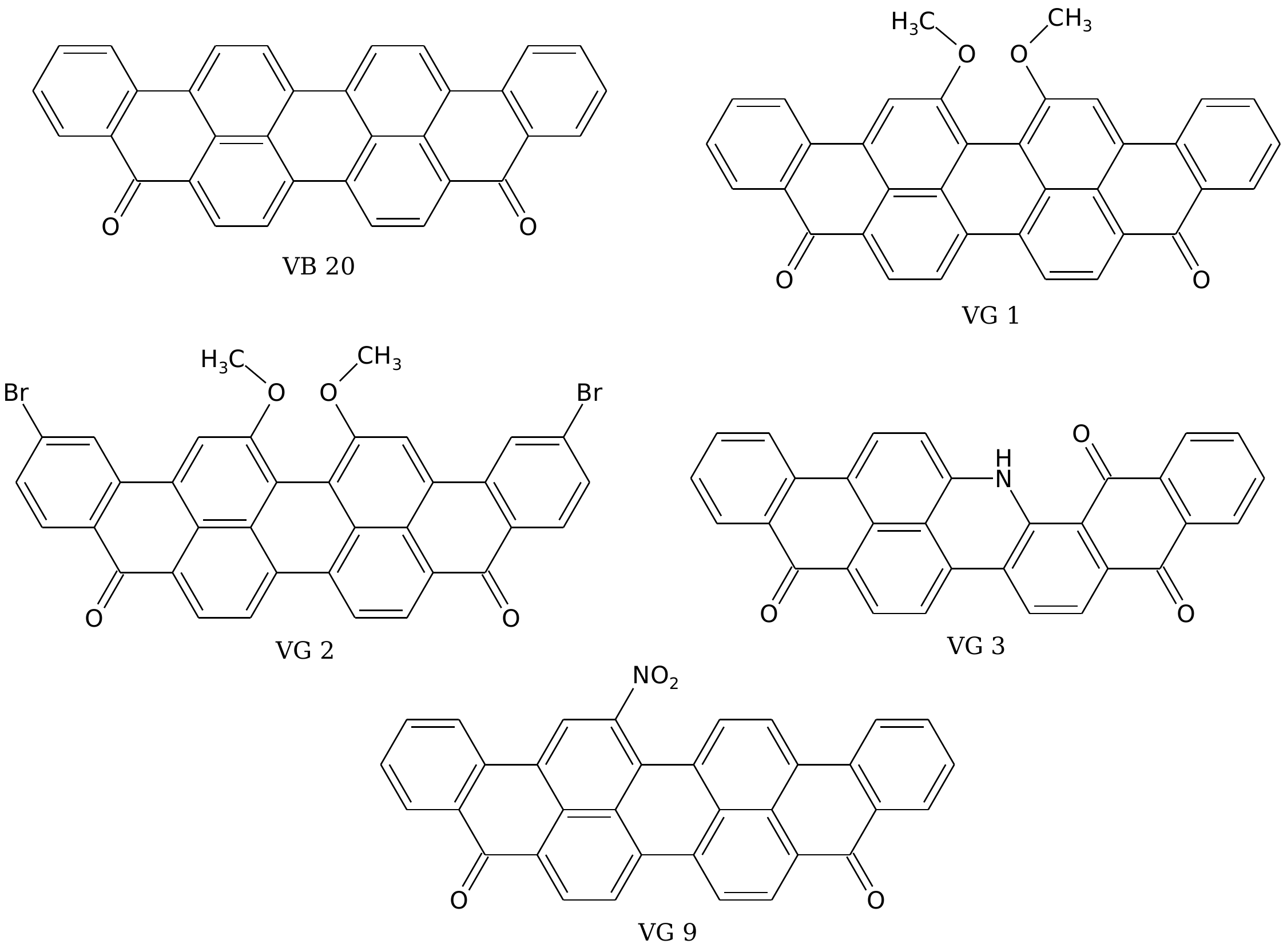}
\caption{Chemical structure of the investigated dyes.}
\label{fig:mol} 
\end{figure}

We are going to use quasiparticle ab initio methods in order to study the excited state properties of materials, which are already known as organic dyes (figure~\ref{fig:mol}): violanthrone (C.I. Vat Blue 20, VB 20) with its derivatives (C.I. Vat Green 1, VG 1; C.I. Vat Green 2, VG 2; C.I. Vat Green 9, VG 9) and the anthraquinone based dye C.I. Vat Green 3 (VG 3). The present work is aimed at evaluating the electronic properties for these materials and checking the suitability of ab initio methods applied here, with respect to a correct description of properties of such materials, which is established by comparison with experimental data.

\vspace{-2mm}
\section{Methods}
\vspace{-1mm}

Firstly, we have performed the structure optimization for all the materials studied here. Secondly, the ground state electron eigenvalues  $ E_{n} $ and eigenfunctions  $ \phi_{n} $ have been obtained from the next equation, using the  generalized gradient approximation (GGA) of the exchange-correlation potential $ V_\text{xc} $ in the form proposed by Perdew, Burke and Ernzerhof \cite{PBE}:
\begin{equation}
[\nabla^2+V_\text{ext}(\textbf{r})+V_\text{H}(\textbf{r})+ V_\text{xc}(\textbf{r})]\phi_{n}(\textbf{r}) = E_{n}(\textbf{r})\phi_{n}(\textbf{r}),
\label{eq:ks}
\end{equation}
where $ \nabla^2 $ is the kinetic energy operator, $ V_\text{ext}(\textbf{r}) $ is the Coulomb potential of the nuclei and core electrons, and exchange-correlation potential of core electrons, $ V_\text{H}(\textbf{r}) $ is the Hartree potential, $ V_\text{xc}(\textbf{r}) $ stands for the exchange-correlation potential, and $ n $ denotes a band index. Thirdly, quasiparticle eigenvalues and eigenfunctions were searched from the following equation: 
\begin{equation}
[\nabla^2+V_\text{ext}(\textbf{r})+V_\text{H}(\textbf{r})]\phi_{n}(\textbf{r}) + \int\Sigma(\textbf{r}, \textbf{r}', E_{n})\phi_{n}(\textbf{r}')\rd\textbf{r}' = E_{n}(\textbf{r})\phi_{n}(\textbf{r}),
\label{eq:gw}
\end{equation}
where $ \Sigma(\textbf{r}, \textbf{r}', E_{n}) $ is the non-local self-energy operator which stands for a renormalized electron states in the many-body system \cite{Onida2002}.
Equation~(\ref{eq:gw}) has been solved within the self-consistent (sc-GW) procedure and with a spherical truncation of the Coulomb term in reciprocal space \cite{PhysRevB.73.233103}.

In the Bethe-Salpeter (BS) formalism, the electron-hole two-body basis can be expressed in the next form
\begin{equation}
L = (H - w)^{-1}F,
\label{eq:bse}
\end{equation}
\begin{equation*}
H =
 \begin{pmatrix}
 R  &  C \\
-C^{*} & -R^{*}
 \end{pmatrix}, 
\end{equation*}
\begin{equation*}
F =
 \begin{pmatrix}
 1  &  0 \\
 0  & -1
 \end{pmatrix}, 
\end{equation*}
where $ L $ is two-particle Green's function,  whose the polarizability can be evaluated with \cite{Onida2002}. The matrix blocks  $ R $ ($ R^{*} $) represent the resonant coupling between electron-hole excitations (recombination), and $ C $ ($ C^{*} $) describe the non-resonant coupling between excitations (recombination). Equation~(\ref{eq:bse}) has been solved by the direct diagonalization procedure.

All calculations have been performed using ABINIT package \cite{abinit}. ABINIT uses a plane wave basis set to compute the electronic density and derived properties. Thus, we determined optimal plane-wave cut-off energies having performed a convergence study that shows the following optimal values: 871~eV (32~Ha) for ground-state runs, 82~eV (3.0~Ha) for the dielectric matrix in random phase approximation (for sc-GW) as well as in BS formalism, 1090~eV (40~Ha) for the exchange part of the self-energy operator. Prior to performing the analysis, we optimized the molecular structures by simple relaxation of ionic positions according to (converged) forces obtained based on the GGA approach. The vacuum spacing is set to be 4~A in each positive or negative direction for all molecules.

\section{Results and discussion}

Electronic properties including the electron energy gap $ E_\text{g} $ calculated within three different approaches, HOMO-level energy $ E_\text{HOMO} $, are presented in table~\ref{tab:results}.

\begin{figure}[!t]
\centering
\includegraphics[width=\linewidth]{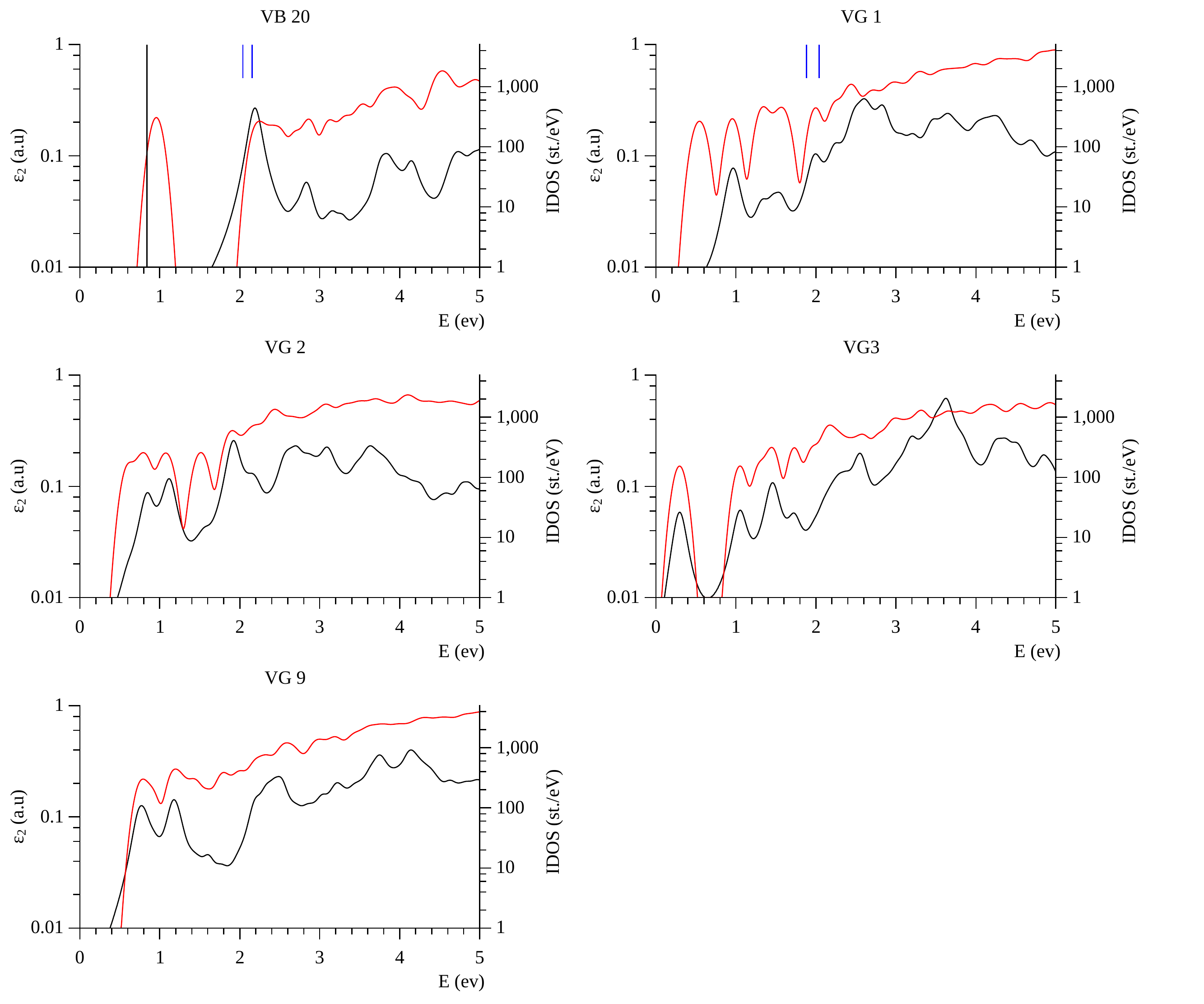}
\caption{(Colour online) Imaginary part of the DF $ \varepsilon_2 $ (black line) and DOS (red line) obtained for molecules. Experimental absorption peaks are indicated by blue lines \cite{doi:10.1021/jo00891a014, B907913K, Liu2010}. Black solid line for VB~20 corresponds to the experimental value of $ E_\text{g} $ \cite{doi:10.1063/1.1700784}. The experimental data correspond to solid-state materials.}
\label{fig:BSE} 
\end{figure}

The values of $E_\text{g}^\text{GGA}$, obtained within the GGA approach, vary from 0.97~eV for VG 3 up to 1.25~eV for VB 20. The sc-GW results differ both in absolute values as well as in relative values. Here, VG 3 has the widest $ E_\text{g}^\text{GW}$, found at the GW level,  which is above 6 times greater than $ E_\text{g}^\text{GGA} $ and equals 6.07~eV.
In case of BS approach, the calculated gap values $ E_\text{g}^\text{BS} $, derived at the BS level,  are lower than $ E_\text{g}^\text{GGA} $ (table~\ref{tab:results}). Similar to the GGA approach, VG 3 has a smaller energy gap $ E_\text{g}^\text{BS} $ (0.95~eV) and VB 20 has the widest gap (0.29~eV).
The obtained real and imaginary part of the BS dielectric function (DF) and interband density of states (DOS) for the list of materials are presented in figure~\ref{fig:BSE}.

Having non-zero DOS around 1~eV, PB 20 is the only molecule from the list which has no DF maximum in IR region. The calculated $ E_\text{g}^\text{BS} $ (table~\ref{tab:results}) agrees with the experimental one which equals 0.84~eV (figure~\ref{fig:BSE}). The location of the first absorption peak coincides with DF maxima (figure~\ref{fig:BSE}).  
The lowest BS gap $ E_\text{g}^\text{BS} $ is obtained in VG3. Here, we can see a DOS maximum, located around 0.3~eV, with a strong DF response (figure~\ref{fig:BSE}). Then, in the range from 0.5 to 0.8~eV, we observe DOS gap. 

\begin{table}[!b]
\caption{The parameters of electronic structure: the band gap $ E_\text{g} $ and the highest occupied molecular orbital energy $ E_\text{HOMO} $.}
\label{tab:results}
\vspace{2ex}
\centering
\begin{tabular}{|c |c| c| c| c|}
\hline\hline
 Molecule & $ E_\text{g}^\text{GGA\strut} $, eV & $ E_\text{g}^\text{GW} $, eV & $ E_\text{g}^\text{BS} $, eV & $ E_\text{HOMO} $, eV\\
\hline\hline
 VB 20 & 1.25  & 5.37 & 0.95 & $-4.81$ \\

 VG 1 &  1.09 & 5.03 & 0.51 & $-5.30$ \\

 VG 2 & 1.10 & 4.91  & 0.60 & $-5.10$ \\

 VG 3 & 0.97 & 6.07  & 0.29 & $-5.94$ \\

 VG 9 & 1.20  & 5.43 & 0.74 & $-6.02$ \\

\hline\hline
\end{tabular}
\end{table}

\section{Conclusions}
In this paper, we have presented the electronic structure and dielectric function $ \varepsilon_2 $ of five polycyclic hydrocarbons based dyes. The ground state properties were calculated within the GGA approximation to the exchange-correlation potential. Excited electronic states were evaluated in order to compare them with the optical absorption spectrum. The primary purpose was to calculate the GW self-energy in order to obtain accurate quasiparticle energies that include the static electron-hole interaction. The secondary purpose was to obtain the BS solutions, including the dynamical electron-hole interaction. The dielectric function obtained by means of the Bethe-Salpeter equation contains information on the energy of exciton excitations that may be compared to measured optical absorption. The results obtained here may be considered as the basis for assessing the validity of the  GGA, GW and BS approaches, and for obtaining theoretical absorption spectrum that would be well compared to the measured one. We found, that only the BS approach provides a good comparison with the data of optical spectroscopic measurements. 
The results for quasiparticle energies, obtained here for finite systems, show a significant difference compared to those found for crystals  \cite{syrotyuk2017quasiparticle, syrotyuk2018calculation}. If for a crystal, parameter $ E_\text{g} $, found in the approach of GW, is close to the experimental value, then for molecules, only the energies of quasiparticle excitations, obtained from BS equation, are well compared with the experimental data.

\ukrainianpart

\title{Екситонні властивості пігментів на основі поліциклічних ароматичних вуглеводнів}
\author{Ю.В. Клиско, С.В. Сиротюк}
\address{
 Кафедра напівпровідникової електроніки, Національний університет ``Львівська політехніка'', \\вул. С. Бандери, 12, 79013 Львів, Україна}

\makeukrtitle

\begin{abstract}
\tolerance=3000
В даній роботі представлено вивчення  пігментів на основі поліциклічних ароматичних вуглеводнів з використанням  ab initio  методів. Список досліджених матеріалів включає промислові зелені пігменти, похідні бензантрону, які є перспективними для використання в органічній електроніці. На першому етапі власні функції та власні значення були отримані з використанням узагальненого градієнтного наближення (GGA). На другому етапі квазічастинкові електронні властивості розраховані в рамках наближення GW. Енергії екситонів, уявна частина діелектричної функції та густина станів були одержані з рівняння Бете-Солпітера (BS). Отримані електронні властивості добре зіставляються з наявними екпериментальними даними.

\keywords органічні напівпровідники, ab initio, електронна  структура

\end{abstract}

\end{document}